\documentstyle[twocolumn,aps]{revtex} 
\begin{document}

\draft

\title{Photofission of Heavy Nuclei at Energies up to 4 GeV}
\author{C. Cetina, B. L. Berman, W. J. Briscoe, P. L. Cole\cite{PLC},\\
G. Feldman, P. Heimberg, L. Y. Murphy, S. A. Philips, J. C. Sanabria\cite{JCS}\\}
\address{{\it Center for Nuclear Studies, Department of Physics}, \\
{\it The George Washington University, Washington, DC 20052}\\}
\author{Hall Crannell, A. Longhi, D. I. Sober\\}
\address{{\it Department of Physics, The Catholic University of America},\\
{\it Washington, DC 20064}\\}
\author{G. Ya. Kezerashvili\cite{GYK} \\}
\address{{\it Budker Institute of Nuclear Physics,
630090 Novosibirsk, Russia}\\}
\date{\today}
\maketitle

\begin{abstract}
Total photofission cross sections for $^{238}$U, $^{235}$U, $^{233}$U, 
$^{237}$Np, $^{232}$Th, and $^{nat}$Pb have been measured {\it
simultaneously}, using tagged photons in the energy range 
E$_{\gamma}$=0.17-3.84 GeV. 
This was the first experiment performed using
the Photon Tagging Facility in Hall B at Jefferson Lab.
Our results show that the photofission cross section for $^{238}$U relative
to that for $^{237}$Np is about $80\%$, 
implying the presence of important processes that compete with
fission.
We also observe that the relative photofission
cross sections do not depend strongly on the incident photon energy
over this entire energy range. 
If we  assume that for $^{237}$Np the photofission probability is equal
to unity, we observe a significant shadowing effect, starting below 1.5 GeV.
\end{abstract}


It has long been assumed \cite{ahr84,fro92} that the photofission cross
section $\sigma_{\gamma F}$ for
the uranium and transuranic isotopes should exhaust the total nuclear photoabsorption cross
section $\sigma_{\gamma A}$ for incident photon energies above 50 MeV, 
so that the fission probability of these nuclei at these energies 
was generally assumed to be unity. 
If this were the case, the photofission method would provide a
very convenient way to determine the total photoabsorption cross
section.
This interesting quantity \cite{ahr85a},
not easily determined by other methods for heavy nuclei, 
would yield information on the intrinsic properties of nucleons inside
the nuclear medium, as well as on the hadronic nature of the photon. 

One might assume the total photoabsorption cross 
section per nucleon $\sigma_{\gamma A} / A$ to be independent of A.
In the $\Delta$-resonance region, this has been shown to be plausible
by various measurements for several complex nuclei ranging from Li to U, 
leading to the concept of the ``universal curve''\cite{ahr85b}. 
For several nuclei,
recent measurements have been performed up to 2.6 GeV \cite{muc99}.
If we define the photofissility $W_F$ of a nucleus as the probability
that this nucleus undergoes fission, then we can express $W_F$ as the
ratio $\sigma_{\gamma F} / \sigma_{\gamma A}$. 
A constant $\sigma_{\gamma A} / A$ implies that, for two
nuclei, the ratio of their photofissilities is given by the ratio of
their photofission cross sections per nucleon $\sigma_{\gamma F} / A$.

The assumption $W_F \simeq 1$ had to be reconsidered after a Novosibirsk group
\cite{ilj92}  reported that the fission probability for $^{237}$Np is about
$30\%$ {\it higher} than that for $^{238}$U  in the
photon energy range 60-240 MeV. 
Subsequent measurements performed at 
the Saskatchewan Accelerator Laboratory (SAL) \cite{san99a}
corroborated these results for the same energy range. 
We now observe this trend to be true up to 4 GeV.

This experiment was the first measurement  performed in Hall B
at the Thomas Jefferson National Accelerator Facility (Jefferson Lab) 
using the photon-tagging facility \cite{sob99}.  
An electron beam is passed through a thin
radiator to produce bremsstrahlung. The residual energy of
an electron after bremsstrahlung emission is deduced from its
measured position in the focal plane of the tagger bending magnet. 
The two-layer focal-plane detector array has an energy resolution of $0.1\%$
provided by the highly segmented E-counter layer and a time resolution
of $\sim$1 ns
for a given T-counter in coincidence with a fission-fragment detector.
By requiring a tight time coincidence between a
T-counter and a corresponding E-counter, the
background from secondary sources in the tagger focal plane was significantly reduced. 

The detector system is able to tag photons with energies from $20\%$
to $95\%$ of the incident electron energy at a single magnetic field
setting.
Thus, with incident electron beam energies of 0.88,
1.71, and 4.04 GeV, we were able to span the photon energy range from
0.17 to 3.84 GeV. 
Using these beam energies allowed for sufficient overlap between 
the energy bites to provide a consistency check 
of the absolute normalization for different data sets.
The difference between the 1.71- and 4.04-GeV data in the overlap region
is about 3$\%$. 
Total fluxes of tagged photons of about $3\cdot 10^7$
photons/s were used during the measurements.

The fission targets studied in this experiment were $^{237}$Np, 
$^{238}$U, $^{235}$U, $^{233}$U, $^{232}$Th, and $^{nat}$Pb. 
Each of these targets consists of a thin layer of $\sim$1 mg/cm$^2$ of 
target material deposited on a backing foil 
(100 $\mu$m aluminum for the actinide isotopes and 25 $\mu$m 
mylar for the lead targets).
In order to increase the number of counts per isotope, we used three
targets for each of the actinides and seven for lead. For the
latter, the effective thickness was further increased by a factor of 1.4
by tilting the targets by $45^\circ$ with respect to the incident photon beam.

One of the fission fragments produced by the interaction of an
incident photon with one of these target nuclei was detected for each
event.
The fission-fragment detectors are novel
low-pressure parallel-plate avalanche detectors (PPADs), which are
described in detail elsewhere \cite{san99b}. 
The anode plane of each of the PPADs consists of a vertical array of 25-$\mu$m
diameter gold-plated tungsten-rhenium wires spaced 1 mm apart. 
The cathode plane, positioned 3 mm from the anode, is a 25-$\mu$m foil of aluminized mylar.
The ionization gas is isobutane.
At our operating parameters (15 Torr and 750 V) these devices are
97.5$\%$ efficient, where the small inefficiency is due to the lack of 
total transparency of the anode wire plane.

Each fission-target foil was mounted on a frame and placed in
front of its own PPAD at a distance of 1.3 cm for the actinide targets
and 0.4 cm for lead.
The acceptance was well defined by circular 4-cm diameter collimators
located between the target foil and the anode plane.
All of these target-detector combinations shared
the same reaction chamber and thus the same gas pressure.
The reaction chamber was placed in the beam line immediately 
downstream of the tagger magnet. 
This arrangement has the important
advantage of enabling us to study all of the
fission targets simultaneously, under exactly the same experimental
conditions.
The trigger for the experiment was given by the OR of all the PPAD
signals. 
At typical event rates of $\sim$150 Hz, the chance of having 
two PPADs in accidental coincidence was less than 0.1$\%$.

Typical ADC and TDC spectra for $^{237}$Np are shown in Fig. \ref{raws}.
The ADC spectrum (Fig. \ref{raws}a) shows that the PPADs
provide  a good separation between $\alpha$ particles and fission fragments. 
A cut has been placed in the lowest part of the valley;
the fission fragments are taken to be the events above this cut.
The time coincidence between a PPAD signal and
the associated electron detected in the tagger focal plane results in a 
 peak in the tagger TDC spectrum, as shown in Fig. \ref{raws}b. 
This peak rests on an extended background of random
coincidences due to events caused by untagged photons and by
radioactive decay products.
The 2-ns RF structure of the electron-beam bunches is evident.
By subtracting the random background from the
coincidence peak, we obtain the yield of real fission-fragment events
(Fig. \ref{raws}c).

The dominant sources of systematic uncertainty are the target thicknesses
\cite{san00} and the photon tagging efficiency. Each contributes less than
5$\%$ and neither depends significantly on the photon energy.
Uncertainties in collimator position and anode wire blockage combine
for an additional 2$\%$.
This leads to an overall systematic uncertainty of 7$\%$ in both the
absolute and relative cross sections.
In the latter the cancellation of the flux uncertainty is compensated by the
additional thickness uncertainty owing to the presence of two targets.

Figure \ref{absx} shows our results for the $^{237}$Np, $^{238}$U, and $^{232}$Th 
absolute photofission cross sections
per nucleon $\sigma_{\gamma F}/A$ as a function of the incident photon energy.  
The error bars reflect statistical uncertainties only.
The previously existing photofission data for these three
isotopes are also shown.
The present data are in good agreement at low energies with the data
from Refs. \cite{ilj92,san99a}. 
Our $^{238}$U data agree well with the average of the Mainz
\cite{fro94} and the Frascati \cite{bia93b} data from
the $\Delta$-resonance region up to about 800 MeV. 
In the same energy region our $^{232}$Th data agree with the Frascati
\cite{bia93a} data.

In Fig. \ref{nppb} we present a comparison between the photofission
data for $^{237}$Np, $^{238}$U, and $^{232}$Th and the existing Pb
photoabsorption data \cite{muc99,cho83,bia96,bro73}. 
The photofission data include the data presented in this paper as
well as previous data from \cite{ilj92,san99a,fro94,bia93b,bia93a}.
Both $\sigma_{\gamma F}$ and $\sigma_{\gamma A}$ data were fitted using
a modified Breit-Wigner formula.
We note that the only low-energy data on Pb
\cite{cho83} should be regarded as a lower limit since it includes
only $(\gamma,xn)$ cross sections, where $x\geq 2$, and hence does not
include any of the $(\gamma,n)$, $(\gamma,p)$, $(\gamma,\alpha)$, 
$(\gamma,pn)$, or $(\gamma,\alpha n)$ channels.
One can see that $\sigma_{\gamma F} /A$ for $^{237}$Np is a few
percent higher than $\sigma_{\gamma A} /A$ for Pb only below the $\Delta$
peak, and is in very good agreement at higher energies.

Using the highest cross section as a reference, 
namely that for $^{237}$Np, we show in Fig. \ref{relx} the photofission cross sections per
nucleon for $^{238}$U and $^{232}$Th relative to that for $^{237}$Np, 
as a function of the incident photon energy.
One can see that the probability for $^{238}$U to undergo fission is
about 20$\%$ {\it smaller}
than that for $^{237}$Np, while $^{232}$Th fissions with about half
the probability of $^{237}$Np. 

The energy dependence of the relative cross sections appears
to be almost flat, suggesting that a
common mechanism is responsible for the photofission process.
This suggests in turn that the two-step
cascade-evaporation model \cite{kik68}, which is used
to explain the fission process at intermediate energies
\cite{ilj97}, is also valid in the 1-4 GeV region. 
According to this model, the incident projectile initiates an
intranuclear cascade.
On a very short time scale (10$^{-22}$ s), 
some of the excited particles exit the nucleus, carrying away both
energy and charge 
(pre-equilibrium emission). The remaining energy is distributed
among the other nucleons to form a residual compound
nucleus in a highly excited state.
After reaching equilibrium, the compound nucleus, over
an extended time period (10$^{-19}$ s), loses
its excitation energy either by evaporating particles (mostly neutrons) 
or by undergoing fission at any stage in the process.

Figure \ref{zza} depicts these relative cross sections as a function of the 
fissility parameter Z$^2$/A.
The vertical bars represent the range spanned by the relative cross
sections over the energy range of the present experiment.
Insofar as the logarithmic dependence of the relative cross section
approaches an asymptotic value, one can infer that the fission
probability approaches unity.
This means that for $^{237}$Np the fission probability, while perhaps
not reaching unity, probably is close to unity.

In the energy range available at Jefferson Lab, photoabsorption
leads to the excitation of bound nucleons and the production of baryon
resonances which behave quite differently  inside the nuclear medium, 
compared with the case for free nucleons.
The D$_{13}$ and F$_{15}$ resonances, clearly seen in the
photoabsorption on the proton and the deuteron \cite{arm72}, 
and still seen in $^3$He \cite{cor96}, are not
observed in the total cross sections for A $\ge$ 4 nuclei.
Photoabsorption measurements have been done for Li, C, Al, Cu, Sn, and Pb
\cite{bia96}, and photofission measurements for Th \cite{bia93a} 
and U \cite{fro94,bia93b}. 
As expected from most previous data, the present photofission cross sections 
 exhibit no prominent resonance structure 
above the $\Delta$ resonance.

Figure \ref{aeff} shows the ratios of the photofission cross sections for $^{237}$Np,
$^{238}$U, and $^{232}$Th and the corresponding cross sections for the
sum of the protons and neutrons \cite{arm72} in each of these nuclei.
Two data points \cite{mic77} for the photoabsorption
cross section for $^{238}$U are shown for comparison.
In order to interpret these data, we recall that
photons with incident energies above the resonance region
begin to exhibit hadronic behavior, and the total cross section starts
to show the onset of the ``shadowing'' effect \cite{wei74,bau78}. 
Since this behavior results in greatly increased interaction strength 
for the photon with hadronic matter, shadowing in nuclei manifests
itself as the total photoabsorption cross section gradually evolves
from purely volume absorption ($\propto$ A) towards purely surface
absorption ($\propto A^{2/3}$).
This effect can be quantified by use of the expression 
$A_{eff}/A=\sigma_{\gamma A}/(Z\sigma_{\gamma p}+N\sigma_{\gamma n})$, 
where $A_{eff}$ represents the
effective number of nucleons seen by the incident photon,
while $\sigma_{\gamma p}$ and $\sigma_{\gamma n}$ are the free-nucleon
photoabsorption cross sections for the proton and the neutron, respectively.
We have seen that $\sigma_{\gamma F}$ is smaller than $\sigma_{\gamma A}$
for the three uranium isotopes, for Th, and for Pb.
Thus, we can not use the photofission
cross section to measure the shadowing effect, except perhaps for the case
of $^{237}$Np (if we assume that its photofission probability is equal
to unity, as suggested by Fig. \ref{nppb}). 
If we make this assumption, we observe the onset of shadowing
below 1.5 GeV, such that $A_{eff}/A$ decreases to $\sim$75$\%$ at 3.6
GeV, as shown in Fig. \ref{aeff}.
It is also clear that $\sigma_{\gamma F}$ for $^{238}$U is much less
than $\sigma_{\gamma A}$ , while $\sigma_{\gamma F}$ for $^{237}$Np is
not.
This trend supports the hypothesis that the fission probability for
$^{237}$Np is close to 100$\%$, and is quite consistent with the
relative fission cross sections of Fig. \ref{zza}, which appear to be
approaching an asymptote at the $Z^2/A$ value of $^{237}$Np.

In summary, we have performed a simultaneous measurement of the total
photofission cross sections for $^{237}$Np, three uranium isotopes,
$^{232}$Th, and $^{nat}$Pb in the photon energy range from 0.17 to 3.84 GeV. 
The relative photofission cross sections show that of these
nuclei, $^{237}$Np has the highest probability to undergo fission, and
that the fission probability for $^{238}$U, which often has been
assumed to be equal to unity, is in fact only about $80\%$ of that of
$^{237}$Np, and does not depend strongly on the incident photon energy up to 4 GeV. 
These results invalidate the use of
the photofission reaction alone to determine the total photoabsorption
cross section for heavy nuclei;
a detailed investigation of {\it all} of the exit channels following
photoabsorption is required to understand completely the microscopic
mechanism governing this process.

This work was supported in part by the U.S. Department of Energy under Grant
No. DE-FG02-95ER40901 and by the National Science Foundation. 
W. R. Dodge, V. G. Nedorezov, and A. S. Sudov contributed during various
stages of preparing the experimental equipment.
We thanks those who took shifts
during the nine days of data taking: D. Branford, B. Carnahan, K. S. Dhuga, 
M. Dugger, J. T. O'Brien, B. G. Ritchie, and I. I. Strakovsky.
We were aided in handling the data-acquisition system by the
expertise of S. P. Barrow.
We would like to acknowledge the contribution of the Jefferson
Lab staff, especially B. A. Mecking and E. S. Smith, for their
advice and support. 
Finally, special thanks are due to G. V. O'Rielly for
constructive discussions during the data analysis.

\begin{figure}
\caption{(a) Typical PPAD ADC spectrum for $^{237}$Np. (b) Typical spectrum for the 
time difference between the detection of a fission fragment and a
tagged electron. The curve
represents an empirical fit to the random background. (c) Same as
(b) after subtraction of accidental coincidences.}
\label{raws}
\end{figure}

\begin{figure}
\caption{Absolute photofission cross sections per nucleon as a function
of the incident photon energy for $^{237}$Np, $^{238}$U, and $^{232}$Th. 
Note that the energy region below 1 GeV has been expanded.
The present results (full circles) are compared with previous
results: \protect\cite{san99a} (open triangles), \protect\cite{ilj92} 
(closed triangles), \protect\cite{fro94}(open circles), and 
\protect\cite{bia93a,bia93b} (stars).}
\label{absx}
\end{figure}

\begin{figure}
\caption{Photofission cross sections per nucleon for $^{237}$Np,
$^{238}$U, and $^{232}$Th (bar bands) and photoabsorption 
cross section per nucleon for Pb (gray band). 
The bands represent fits to the existing data, present
and previous, for Np \protect\cite{ilj92,san99a}, 
U \protect\cite{ilj92,san99a,fro94,bia93b}, Th
\protect\cite{san99a,bia93a}, and Pb \protect\cite{muc99,cho83,bia96,bro73} (see text).
The band widths represent the uncertainties in the fits.
The energy scale is logarithmic, for clarity at the lower energies.}
\label{nppb}
\end{figure}

\begin{figure}
\caption{Photofission cross sections per nucleon relative to $^{237}$Np
as a function of the incident photon energy for $^{238}$U and
$^{232}$Th.
The energy region below 1 GeV has been expanded.}
\label{relx}
\end{figure}

\begin{figure}
\caption{Photofission cross sections per nucleon relative to $^{237}$Np
as a function of the fissility parameter Z$^2$/A.
The vertical bars represent the range spanned by the relative cross
sections over the energy range of the present experiment.}
\label{zza}
\end{figure}

\begin{figure}
\caption{Ratio of the photofission cross sections for $^{237}$Np,
$^{238}$U, and $^{232}$Th and the corresponding total cross sections for the
sum of the protons and neutrons \protect\cite{arm72} in each of these nuclei.
Two data points (closed squares) \protect\cite{mic77} for the total photoabsorption
cross section for $^{238}$U are shown for comparison.}
\label{aeff}
\end{figure}

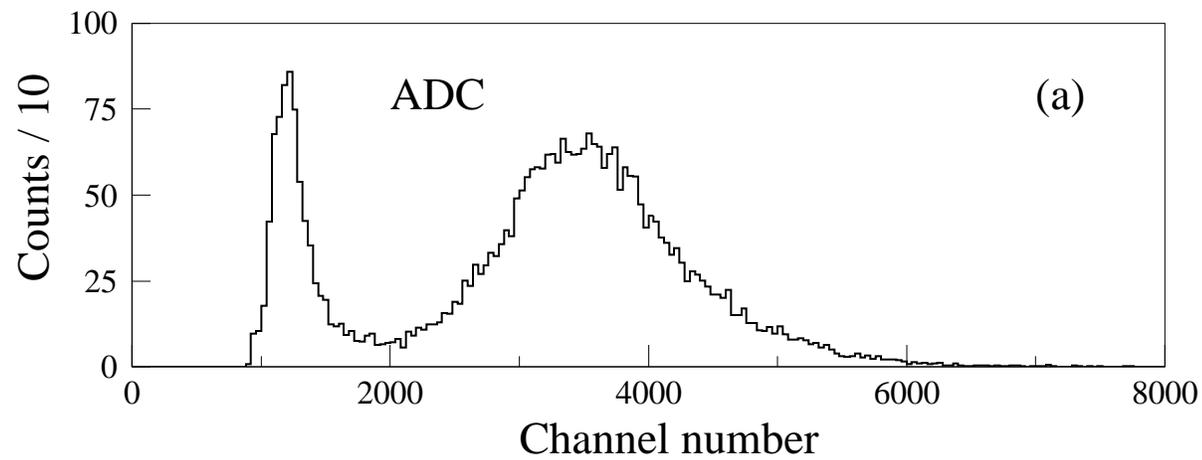
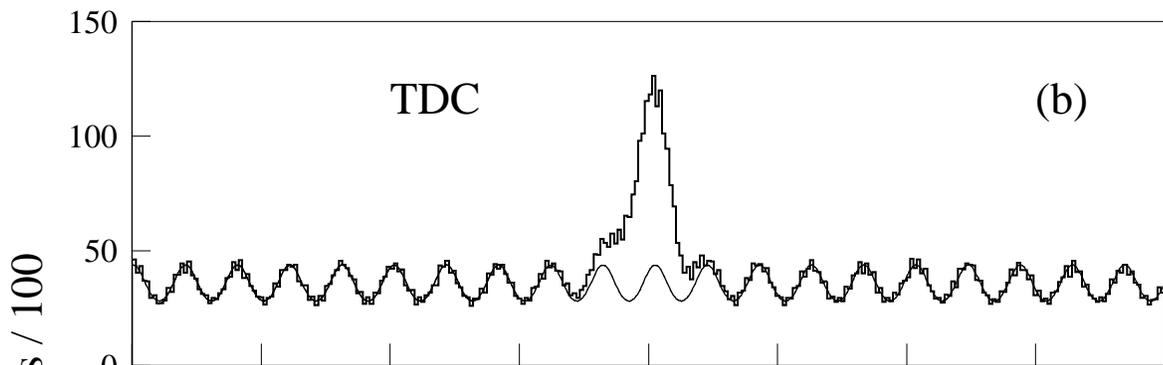
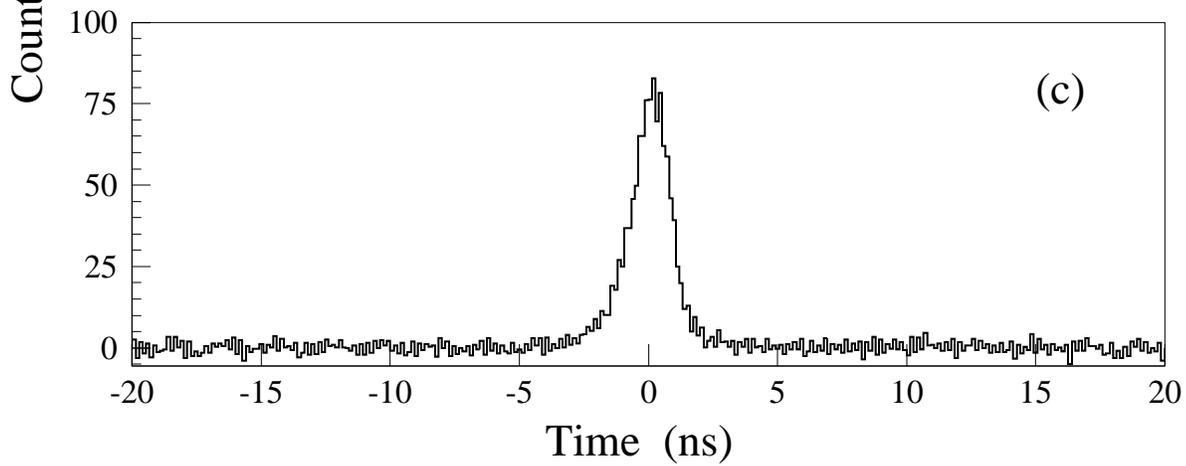
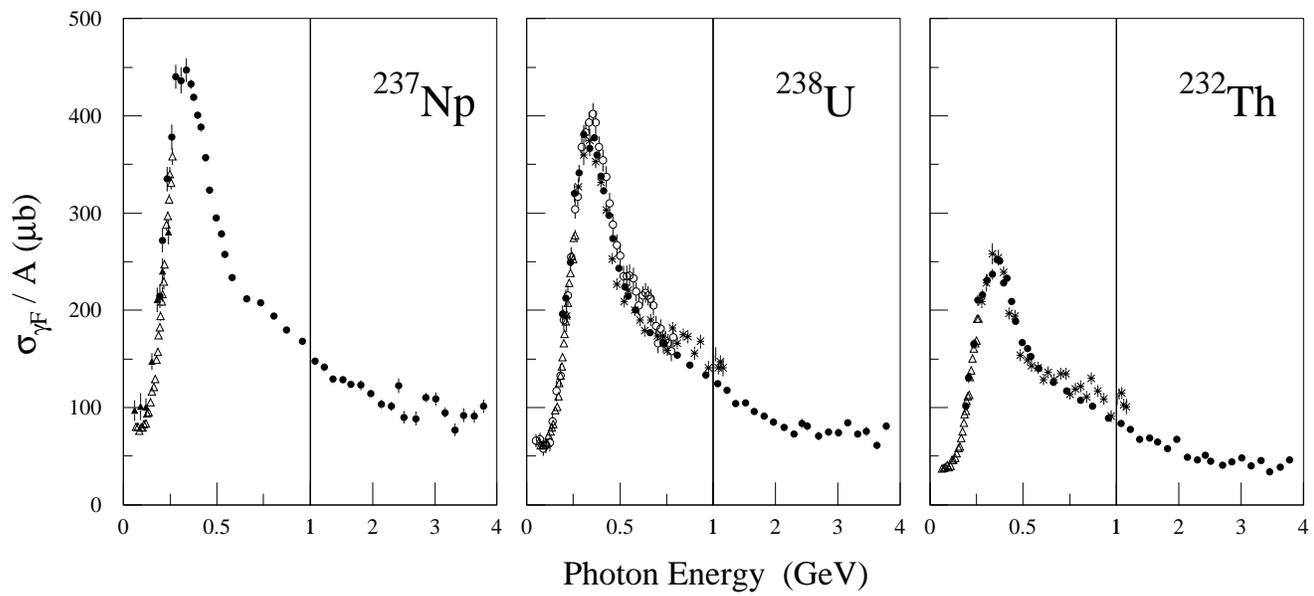
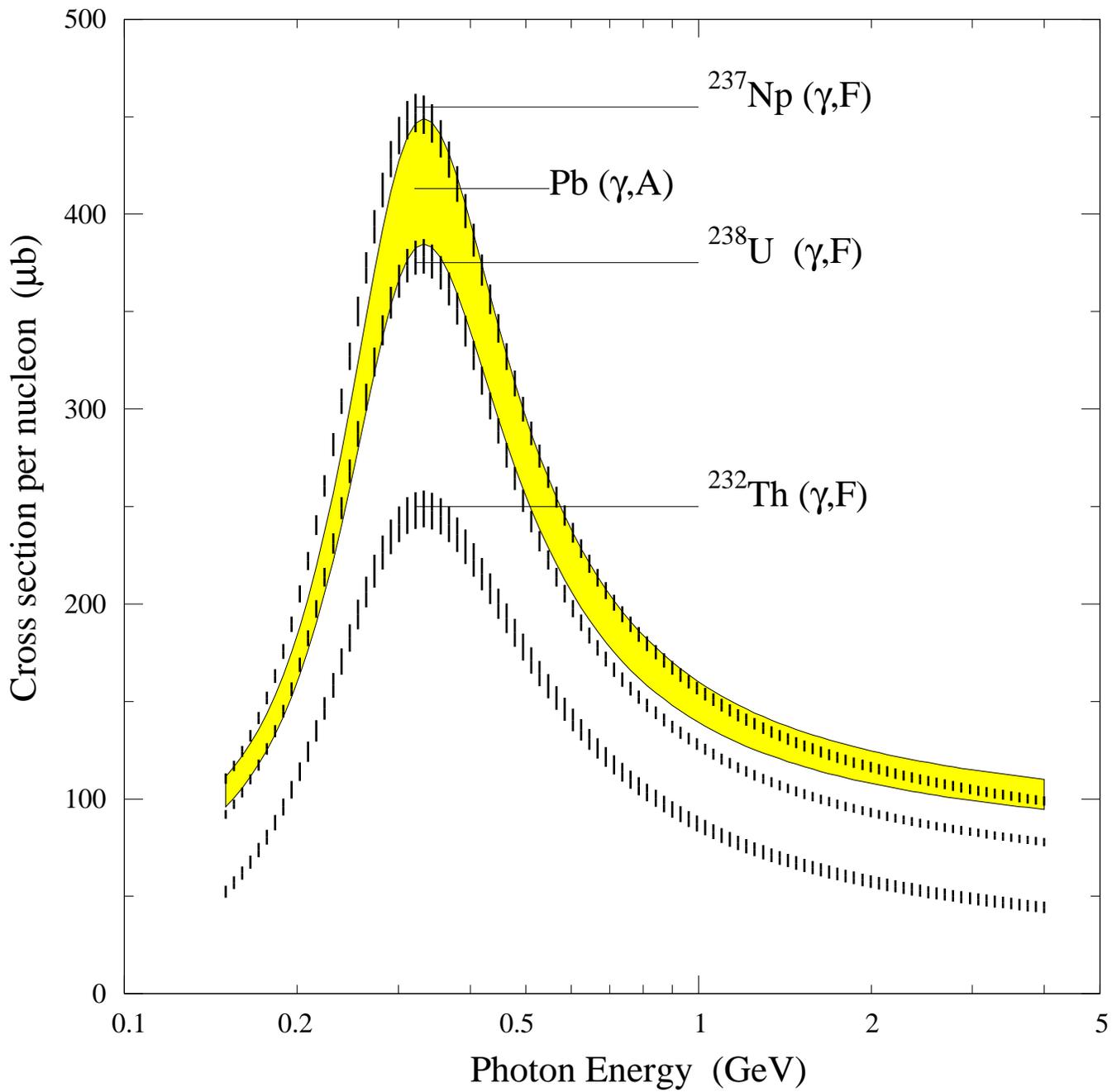
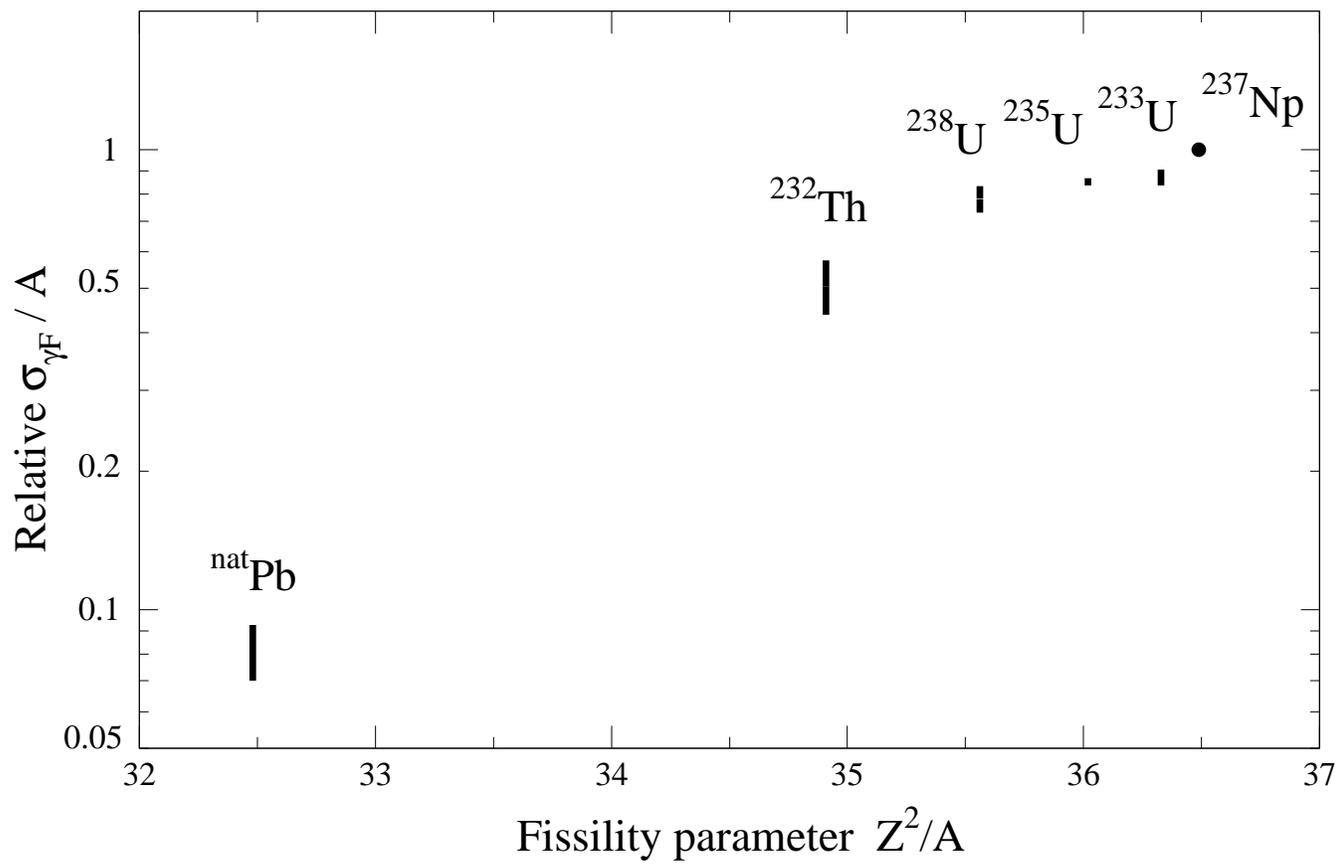
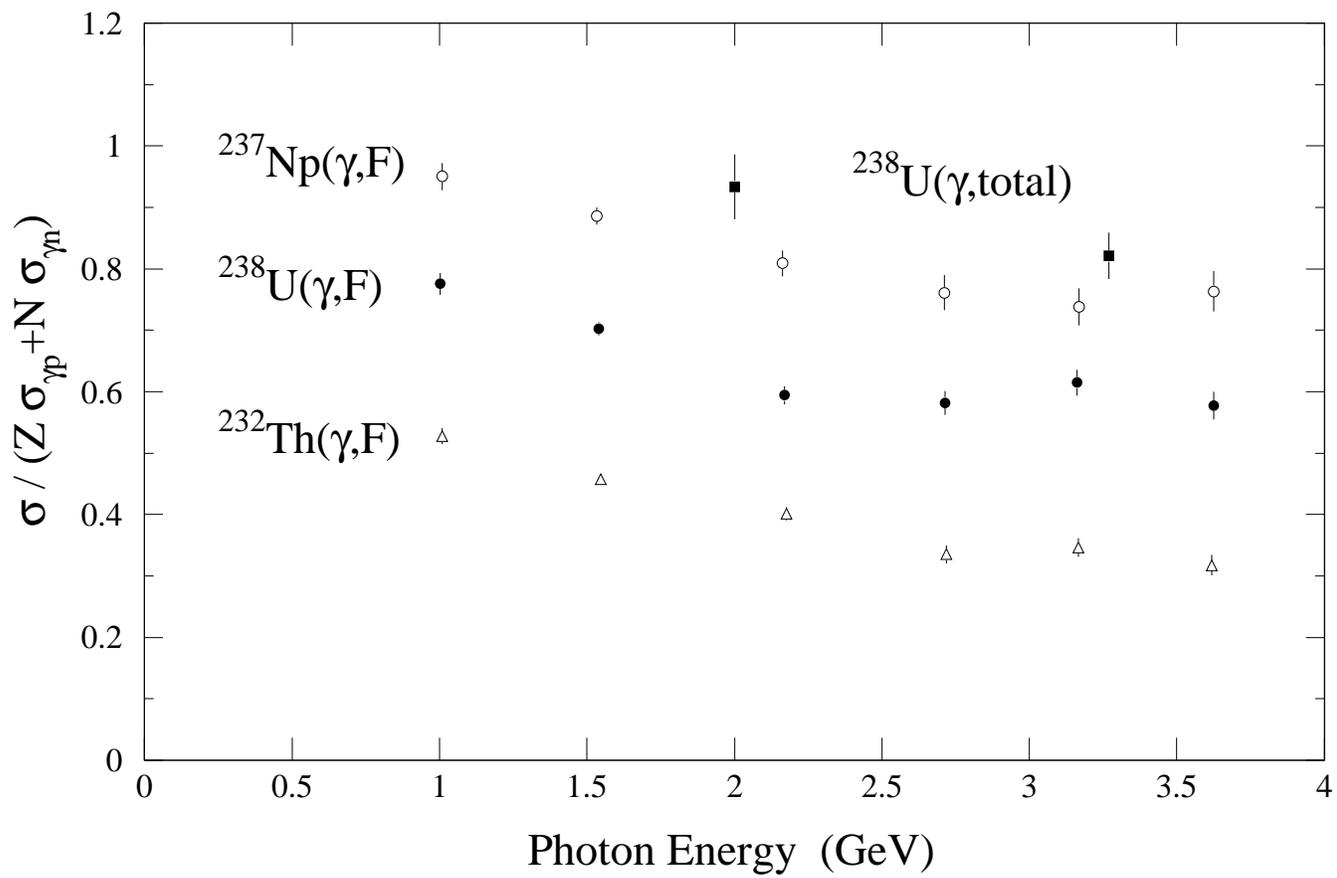
\end{document}